\documentclass[aps,twocolumn,groupedaddress]{revtex4}
\usepackage{epsfig}
\usepackage{graphicx}
\usepackage[T1]{fontenc}
\usepackage{ae}
\usepackage{color}
\usepackage[latin1]{inputenc}
\usepackage{amssymb,amsbsy,amsmath}
\usepackage{bbm}

\begin{document}



\title[Einstein's box]
{Einstein's photon box revisited}

\author{Heinz-J\"urgen Schmidt$^1$
}
\address{$^1$  Universit\"at Osnabr\"uck,
Fachbereich Physik,
 D - 49069 Osnabr\"uck, Germany}


\begin{abstract}
We present a reformulation of Bohr's analysis of Einstein's thought experiment with the photon box, which hopefully settles some unclear issues. The box is weighed by two position measurements, the first immediately before the emission of the photon and the second after a quarter period of oscillation of the box, or alternatively after a half period. These times are precisely defined with respect to the moving clock installed in the box, but vary statistically when measured with a stationary clock outside the box, due to gravitational time dilation in accordance with Bohr's original idea.
Moreover, we will discuss two other proposals to solve the apparent inconsistency without invoking the theory of general relativity.
\end{abstract}

\maketitle

\section{Introduction}\label{sec:I}
Among the most famous thought experiments are the experiments on the consistency check of Heisenberg's uncertainty principle (HUP)
proposed in the Einstein-Bohr debate.
It is the aim of the present note to reformulate Bohr's response to Einstein's photon box thought experiment at the Solvay conference 1930
\cite{B49}
and to discuss other possible responses and related proposals.
The focus is on the physical arguments and not on the historical context.
The theoretical framework of the Einstein-Bohr-debate is {\em not} quantum mechanics
but classical statistical physics augmented by the postulate that only statistical ensembles are admitted that satisfy the HUP.
We will refer to this modified classical theory as "semi-classical"
without using this term in a technical sense of "first order in $\hbar$" or the like.
The Einstein-Bohr debate (at least its may be un-historical reconstruction considered here) is about
the consistency of such a theory: Is it possible to construct a measurement scheme within this theory that disproves the HUP?
Einstein's photon box thought experiment is hence viewed as a proposal to prepare an ensemble of photons such that the spread of the
emission times $\delta t$ and the spread of the energies $\delta E$ can both be made arbitrarily small. Bohr's response will be
reconstructed as a detailed analysis of the proposed preparation and measurement scheme showing that, on the contrary, the HUP
of the form $\delta t\, \delta E\gtrsim h$ is nevertheless satisfied. Typically, the analysis need not be too precise; it would
suffice to obtain order of magnitude estimates corresponding to the vagueness of the statement $\delta t\, \delta E\gtrsim h$,
which should be read as ``the product of $\delta t$ and $\delta E$ is bounded from below by a non-zero quantity of order of magnitude $h$".

While Bohr's general line of defense, including appeal to gravitational time dilation, is well known,
it is difficult (for me) to follow the details of the arguments usually put forward in the literature,
and therefore a careful reconstruction seems to be appropriate.
This also has a pedagogical aspect: I think the treatment of Einstein's thought
experiments lends itself very well to lectures because various elementary physical arguments are interwoven;
from there, a consistent presentation might be useful.

The paper is organized as follows. In Section \ref{sec:G} we present the general setting of the photon box experiment
and consider particular versions which we find easier to analyze. After the emission of the photon the box will oscillate
about its new rest position with the period $T$.
The second weighing can be replaced by a measurement of the amplitude of the oscillation, see Section \ref{sec:TV}, i.~e., by a position
measurement either after the time $T/4$, see subsection \ref{sec:FV}, or after the time $T=T/2$, see subsection \ref{sec:SV}.
In both cases, the energy-time HUP can be derived from the fact that transformation between coordinate time and proper time varies
according to uncertainty of the exact motion of the box.
It has been noted in the literature that Bohr could have refuted Einstein's photon box objection without recourse to general relativity
\cite{T71,BT88,B89}.
We treat this issue in section \ref{sec:EOT},
where we discuss the uncertainty relation devised by P.~Busch that holds for energy uncertainty and the opening time of the shutter
and a similar estimate based on semiclassical quantum theory. However, the shutter is not decisive; another third version
of Einstein's photon box without shutter can also be shown to satisfy the HUP without the need to employ general relativity,
see Section \ref{sec:TV}.
But interestingly, one has to argue with gravitational redshift to explain why the photon increases the weight of the box,
see section \ref{sec:CC}. We conclude with a Summary in section \ref{sec:SUM}.

\section{Generalities}\label{sec:G}

As mentioned in the Introduction the conceptual framework of the Einstein-Bohr discussion of the counterexamples is a modified classical statistical mechanics, where the modification consists in the fact that only statistical ensembles are admitted which fulfill the HUP.
(The treatment of photons is a different topic, but can be left aside here since we will only consider the photon box with the exception of Section \ref{sec:CC}.)
This can be understood as a kind of compromise between the two adversaries: If Bohr insisted on using full quantum mechanics for the discussion, Einstein could complain of a {\em petitio principii}, since his counterexamples, after all, question the very consistency of quantum mechanics. Conversely, modified statistical mechanics as a limiting case of quantum mechanics should be acceptable to Bohr, even more so if it is already sufficient to invalidate the counterexamples. Statistical mechanics without HUP, on the other hand, would be unacceptable to Bohr, and also uninteresting for the discussion. The interpretation of the uncertainty thus becomes ambiguous, depending on whether one understands it purely classically or quantum mechanically, but this is not a shortcoming, but a feature of the discussion.

\begin{figure}[t]
\centering
\includegraphics[width=1.0\linewidth]{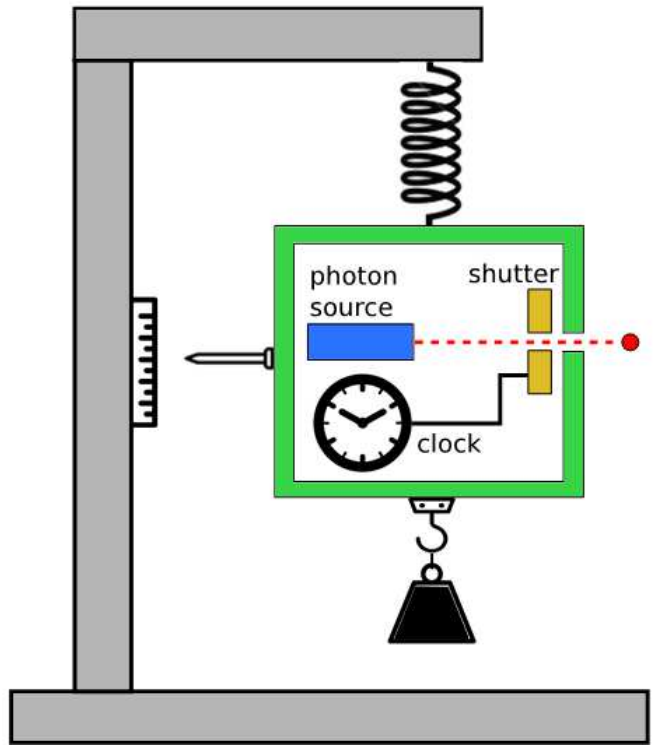}
	\caption{A schematic drawing  of the photon box, from \cite{EPB}.}
\label{FIGBOX}
\end{figure}

Einstein's thought experiment consists, in short, in the emission of a photon from a photon box at an arbitrarily sharply defined time and a determination of its energy by a weighing of the box before and after the emission, invoking the famous $E=m\,c^2$. Bohr's reply rests on an analysis of the weighing process.
The box is suspended from a spring in a gravitational field and its weight is determined by reading the vertical position of the box on a pointer, see Figure \ref{FIGBOX}.
As has been noted several times in the literature \cite{T71,TDG00}, Bohr's original arguments are difficult to follow. By analogy with a balance the weighing is performed by attaching smaller and smaller weighing pieces until the position of the pointer on the box returns to the original position, except for a residual uncertainty
$\delta z$. It is not entirely clear whether one waits for complete damping of the box' oscillations, which may involve further energy dissipation that would be difficult to be determined. Moreover, the Bohr weighing process seems to amount to very many position measurements, classically harmless, but quantum mechanically questionable, because additional perturbations of the photon box system would have to be taken into account.
Finally, the decisive argument is as follows: If the final measurement phase takes time $\widetilde{T}$, then the clock in the photon box accumulates a time uncertainty $\delta \widetilde{T}$ due to gravitational time dilation and position uncertainty $\delta z$ in the gravitational field.
But why does $\delta \widetilde{T}$ coincide with the uncertainty $\delta t$ of the emission timing?

To cope with these problems, we will modify the weighing procedure by neglecting damping and calculating the mass loss $m$  from
the amplitude of the oscillation of the box after the emission of the photon.
Thus
we measure the position of the box pointer just before the time $\tau=0$ of emission and then at a fixed time $\tau=T/4$ (or $T/2$) after emission.
(The time values denoted by $\tau$ always refer to the clock installed in the box and $T$ is the period of the oscillation of the box).
From the difference of the two position measurements we determine the amplitude of the oscillation of the box and from this the mass loss $m$
corresponding to the photon's energy $E=m\,c^2$.
At the time immediately after the emission the system is described by a probability distribution in the phase space which satisfies the HUP. After that a classical time evolution with a harmonic oscillator Hamiltonian is assumed, where also the classical uncertainties evolve in such a way that according to Liouville's theorem the HUP holds unchanged. The final uncertainties $\delta E$ and $\delta t$ are interpreted as statistical fluctuations (standard deviations) in an ensemble realized by a large number of repetitions of the single experiment. (Admittedly, this is not the only possible interpretation of the HUP, see the Summary.) Thus, the Einstein photon box experiment (in the present modified version) is understood as a preparation procedure for an ensemble of photons in which both time measurements and energy measurements can be made (but not simultaneously in a single experiment) with the results $\delta E$ and $\delta t$ for the statistical fluctuations. We do not have to care about the corresponding quantum mechanical observables (generalized observables in the case of time measurements) since we only do (semi-)classical calculations.

We now consider $N$ measurements of the arrival times of the photons on a screen or detector in the single experiments as the readings $t^{(1)}, t^{(2)},\ldots, t^{(N)}$ of a stationary ``standard clock". It is clear that these arrival times alone are not sufficient to determine $\delta t$.
They have to be referred to ``time zero points" for each single experiment. Let $t_0^{(1)}, t_0^{(2)},\ldots,t_0^{(N)}$ be these time zero points
(again defined as readings of a standard clock), then the statistics of arrival times are calculated from the differences $t^{(i)} -t_0^{(i)}, i=1, ...,N$.
For example, the zero points could be defined by signals sent to the standard clock from the apparatus in each single experiment.
For the analogous results of energy measurements $E^{(i)}$ the same applies in principle; however, for theoretical reasons one can assume that the results do not depend on time and the specification of $t_0^{(i)}$ will be superfluous.
Of course, we cannot accurately calculate the standard deviations $\delta E$ and $\delta t$ based on the previous data. However, we can give estimates determined by the uncertainty of the measurements at the photon box. For example, it is very plausible that statistical fluctuations $\delta m$ from the results of the two weighings (multiplied by the factor $c^2$) is smaller than the standard deviation $\delta E$ of the photon energy measurement.

The analogous analysis must be performed for the estimation of the statistical fluctuation $\delta t$ of the arrival times of the photons from the single experiments. In the single experiment, the photon is emitted at time $\tau=0$. The exact time is uncertain due to the finite opening time of the shutter. However, it is asumed in the following that this opening time can be chosen arbitrarily small (but see the discussion in Section \ref{sec:EOT}).
As noted above, it is also necessary to define a time zero for the single experiment and to send it to the standard clock. For this, the emitted photon itself cannot be used.  To send a second photon at time $\tau=0$ would be possible in principle, but would unduly complicate the analysis.
We decide to define the time zero point by the very signal that is send from the box to an external experimenter in order to trigger the second position measurement at time $\tau=T/4$ (or $T/2$). The back reaction of this signal (emission of a second photon) on the box is negligible with respect to the second position measurement, because of the arbitrarily small time difference between these two events.

Thus the problem narrows down to the question why the time difference between $\tau=0$ and $\tau=T/4$ (or $T/2$) can vary statistically in such a way that this variation does not become arbitrarily small without influencing other quantities. According to Bohr, the reason is the difference between the readings of the stationary standard clock and the readings of the box clock due to gravitational time dilation.

\section{Two versions of Einstein's box}\label{sec:TV}

As indicated in the preceding Section the weighing process will be performed as follows: The box with rest mass $M+m$, of which a section is shown in Figure \ref{FIGBOX}
is suspended in a spring-balance with spring constant $k$ and is furnished with a pointer to read its position on a scale fixed to the balance support.
The $z$-axis, which represents the vertical scale, is chosen to point upwards and its zero point
corresponds to the rest length of the spring,
so that at the beginning the position of the pointer $z_0<0$ satisfies an equilibrium condition of the simple form
\begin{equation}\label{M1}
 - k\,z_0= (M+m)\,g
  \;,
\end{equation}
where $g$ is the gravitational acceleration, assumed to be approximately constant. After the emission of the photon with energy $E$
into the direction chosen as the $x$-axis the rest mass of the box will be reduced by the amount
\begin{equation}\label{M2}
  m = \frac{E}{c^2}
  \;.
\end{equation}
This follows from special relativity by expanding the Minkowski length of the energy-momentum $4$-vector
\begin{equation}\label{M3}
\left(
\begin{array}{c}
M\,c^2\\
0\\
0\\
0\\
\end{array}
\right)+
\left(
\begin{array}{c}
E\\
E\\
0\\
0\\
\end{array}
\right)
\end{equation}
of the box after the emission into a Taylor series and taking the lowest linear order w.~r.~t.~$\frac{m}{M}$.
Consequently, we will assume
\begin{equation}\label{Mapp}
  m\ll M
\end{equation}
throughout this paper. Although it is unrealistic to assume that the photon will be emitted exactly into a horizontal direction,
the consideration of a vertical recoil of the box would only give rise to a further uncertainty proportional to
$\frac{m}{M}$ and can be neglected.

The reduced mass $M$ of the box after the emission of the photon corresponds to a new equilibrium position $z_1<0$ of the pointer defined by
\begin{equation}\label{M4}
 - k\,z_1= M\,g
  \;,
\end{equation}
such that by (\ref{M1}) we have
\begin{equation}\label{M4}
  k\,(z_1-z_0)= m\,g
  \;.
\end{equation}
The box will perform vertical damped harmonic oscillations about the new equilibrium position $z_1$ until it reaches this position after long times.
A measurement of $z_1$ after long times would then indirectly give the value of $m$ and hence of $E$. However, it would be necessary
to consider the dissipated energy and its mass equivalent to correct the asymptotic equilibrium position $z_1$.

Here we will follow a different strategy based on the observation that the amplitude of the (undamped) oscillation is $z_1-z_0$,
the starting position being $z_0$, and hence
the position $z_1$ will be reached after a quarter period $\frac{T}{4}$ of the oscillation.
Alternatively, the amplitude can be calculated from the position $z_2=2\,z_1-z_0$ of the pointer after a half period $\frac{T}{2}$.
We may use the description of the oscillation according to classical mechanics and thus set
\begin{equation}\label{M5}
 T=\frac{2\,\pi}{\omega},\quad \mbox{where } \omega =\sqrt{\frac{k}{M}}
 \;.
\end{equation}
In this way we will realize the second weighing of the box by a position measurement of the pointer
at the time $\frac{T}{4}$ (first version) or $\frac{T}{2}$ (second version) after the emission of the photon.

\subsection{First version of Einstein's box}\label{sec:FV}

Now we consider the reconstruction of Bohr's answer corresponding to the first version. He may use the condition that all quantities used
do not have sharp values but vary according to the HUP, of course except the quantities $t$ and $E$  under discussion.
Usually it will not be necessary to consider all possible uncertainties but to concentrate on the relevant ones that are
required to obtain $\delta t\, \delta E\gtrsim h$. The other quantities can be treated as if they would have a sharp value.
For example, the first weighing of the box can be performed by means of a position measurement leading to $z_0$
with a corresponding uncertainty for $M$  but this will be irrelevant for what follows (but see the Sections \ref{sec:EOT} and \ref{sec:TV}).

Obvious candidates for relevant uncertainties are the vertical coordinates $z_i$ and the corresponding vertical momenta $p_i$ of the box,
where the index $i=0,1$ refers to the time of emission and of the second position measurement, resp.~. These uncertainties
can be legitimately assumed to satisfy the HUP:
\begin{eqnarray}\label{M6a}
 \delta z_0\, \delta p_0&\gtrsim& h\;,\\
 \label{M6b}
 \delta z_1\, \delta p_1&\gtrsim& h
 \;.
\end{eqnarray}

\begin{figure}[t]
\centering
\includegraphics[width=1.0\linewidth]{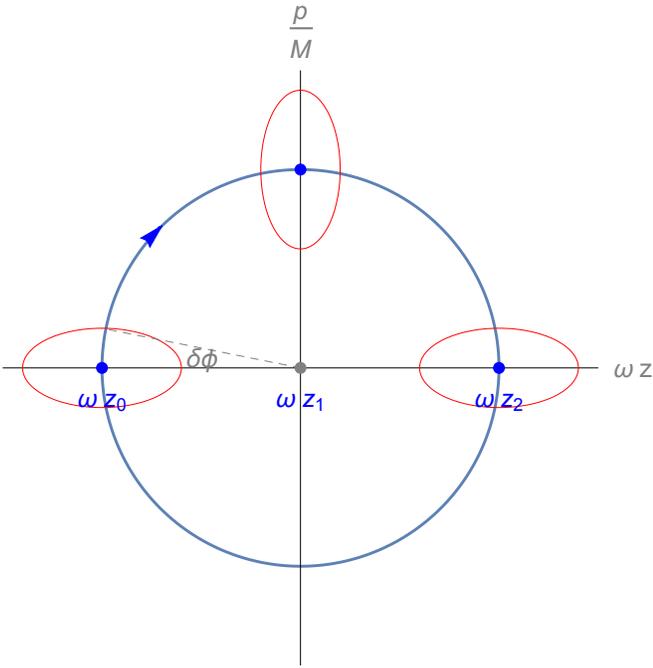}
	\caption{The oscillation of the photon box can be viewed as a rigid rotation in the phase space
with scaled coordinates $\omega\,z$ and $\frac{p}{M}$.
Three events corresponding to the emission of the photon at $\tau=0,\,z=z_0$ and the second position measurement
at $\tau=T/4,\,z=z_1$ or $\tau=T/2,\,z=z_2$ are represented as blue dots on the circle with center $(\omega\, z=\omega\, z_1, p/M=0)$ and radius
$r=\omega\,(z_1-z_0)$.
The uncertainties of these phase space points are indicated by small red ellipses which also rigidly rotate in phase space.
The first ellipse on the far left corresponds to the uncertainty $\delta\phi$ of the phase angle $\phi$
such that $\sin \delta\phi \approx \frac{\delta p_0/M}{\omega(z_1-z_0)}$.
The figure corresponds to the first and the second version of Einstein's box considered in this paper.
}
\label{FIGCIRC}
\end{figure}

Since the oscillation of the photon box can be viewed as a rigid rotation in the phase space
with scaled coordinates $\omega\,z$ and $\frac{p}{M}$, see Figure \ref{FIGCIRC},  the role of the uncertainties is exchanged after a
$90^\circ$ rotation corresponding to the time evolution between the two mentioned instants of time. More precisely, we conclude
without further calculation
\begin{eqnarray}
\label{M7a}
  \omega\,\delta z_1 &=& \frac{1}{M}\delta p_0\;, \\
  \label{M7b}
  \frac{1}{M}\delta p_1 &=&  \omega\,\delta z_0
  \;,
\end{eqnarray}
in accordance with (\ref{M6a}) and (\ref{M6b}). This yields a first estimate of the uncertainty of $m$:
\begin{eqnarray}
\label{M8a}
  \delta m &\stackrel{(\ref{M4})}{\approx}& \frac{k}{g}\,\delta z_1\stackrel{(\ref{M5})}{=}\frac{M\omega^2}{g}\,\delta z_1 \\
  \label{M8b}
   &\stackrel{(\ref{M7a})}{=}&\frac{M\omega^2}{M\omega g}\,\delta p_0= \frac{\omega}{g}\, \delta p_0
   \;.
\end{eqnarray}

Next we come to the uncertainty $\delta t$ of the quantity $t_1$. The latter means the time of the emission of the photon as it is measured
by a stationary clock. This may be different to the time measured by the moving clock inside the box due to
gravitational and special-relativistic time dilatation effects.
To choose a particular physical model we consider the Schwarzschild metric, see, e.~g., \cite{W84}, and its linear approximation in the
vicinity of the photon box. Then the stationary clock mentioned above can be chosen as a clock at a large distance from the center
and hence measuring the Schwarzschild coordinate $t$ to an arbitrary precision. This clock should be used to measure the two
times of emission of the photon at $\tau=0$ and of the second position measurement at $\tau=T/4$ and its difference $t_1$.
This has to be distinguished from the {\em proper time} difference $\tau_1=T/4$ measured by the moving clock installed in the box.
One can, of course, calculate the transformation between $\tau_1$ and $t_1$ if the exact motion of the box is known.
And here the uncertainty $\delta z_0$ comes into play. Different motions of the clock with an initial position $z_0$ varying with
a latitude of $\delta z_0$ lead to different values of the coordinate time difference $t_1$ between emission and second position measurement,
although the proper time between the two events measured by the moving clock has always the same value $\tau_1=\frac{T}{4}$.
This yields the desired value of $\delta t$ that, hopefully, would satisfy the HUP.

Anticipating the result of the following calculation to be
\begin{equation}\label{M9}
  \delta t \gtrsim \frac{g}{c^2\,\omega}\,\delta z_0
  \;,
\end{equation}
we conclude the desired result
\begin{equation}\label{M10}
  \delta t\,\delta E \stackrel{(\ref{M8b},\ref{M9},\ref{M2})}{\gtrsim}
  \left( \frac{g}{c^2 \omega}\delta z_0\right)\,\left(\frac{c^2\omega}{g}\delta p_0 \right)\
  \stackrel{(\ref{M6a})}{\gtrsim} h
  \;.
\end{equation}

In order to derive (\ref{M9}) we consider the Schwarzschild metric \cite{W84}
\begin{equation}\label{M11}
 d\tau^2= \varphi(r) \,dt^2- \frac{1}{c^2 \varphi(r)}\,dr^2
 \;,
\end{equation}
where
\begin{equation}\label{M12}
 \varphi(r) =1-\frac{2\, G\, M_g}{c^2\,r}
 \;,
\end{equation}
$G$ denoting the gravitational constant and $M_g$, e.~g., the earth's mass if the thought experiment is performed in a terrestrial laboratory

Let $r_0$ be the Schwarzschild radial coordinate of the second equilibrium position corresponding to $z_1$
such that
\begin{equation}\label{M13}
 g=\frac{M_g\,G}{r_0^2}
 \;.
\end{equation}
$r_0$ is approximately the earth's radius.
Solving (\ref{M11})
for $\frac{dt}{d\tau}$ yields
\begin{eqnarray}\nonumber
 \frac{dt}{d\tau}&=&\sqrt{\frac{1}{\varphi(\tau)}+ \frac{1}{\varphi(\tau)^2}\frac{v^2}{c^2}}=\varphi(\tau)^{-1/2}+\varphi(\tau)^{-3/2}\frac{v^2}{2\,c^2}\\
\label{M14}
&& +O\left(\frac{v^4}{c^4}\right)
 \;.
\end{eqnarray}
Here we have set  $v=\frac{dr(\tau)}{d\tau}$, assuming  $v\ll c$,  for the vertical velocity of the box
and abbreviated $\varphi(r(\tau))$ by $\varphi(\tau)$.
The series expansion (\ref{M14}) clearly shows that the time dilatation has two components:
The pure gravitational part $\varphi(\tau)^{-1/2}$ and the special-relativistic part $\varphi(\tau)^{-3/2}\frac{v^2}{2\,c^2}$.
For the calculation of $\delta t$ the latter leads to a similar result as the former one but with a factor $\frac{\pi\,m}{M}$ smaller
and hence can be neglected.

Next we will write $\varphi(\tau)=\varphi(r_0+z(\tau))$ where $z(\tau)$ represents the oscillation of the box after the emission of the photon.
A convenient approximation to the correct solution of the equation of motion that would result in general relativity
is an oscillation proportional to $ \cos ( \omega \tau )$, as follows from classical mechanics,
except for the substitution of the time coordinate for the proper time $\tau$.
The amplitude of the oscillation will be written as $z_1-z_0+\delta z_0$ corresponding to
an initial position of the box at $z=z_0-\delta z_0$.
This leads to
\begin{equation}\label{M13a}
r(\tau)= r_0+z(\tau)= r_0+\left(z_0-z_1-\delta z_0 \right) \cos ( \omega \tau  )
\;,
\end{equation}
taking into account the conditions $r(T/4)= r_0$  and $r(0)= r_0+\left(z_0-z_1-\delta z_0 \right)$.
Then
the gravitational time dilation factor $\varphi(\tau)^{-1/2}$ can be further approximated by
\begin{eqnarray}\label{M15a}
 \varphi(\tau)^{-1/2}&=&\frac{1}{\sqrt{1-\frac{2 g r_0^2}{c^2 \left(r_0+z(\tau)\right)}}}\\
 \label{M15b}
 &\approx& 1+\frac{g r_0}{c^2 \left(1+z(\tau)/r_0\right)}\\
 \label{M15c}
 &\approx& 1+\frac{g}{c^2}(r_0-z(\tau))
 \;,
\end{eqnarray}
using that $z(\tau)\ll r_0$ in (\ref{M15c}) and $\frac{2\,g\,r_0}{c^2}=\frac{2\,G\,M_g}{c^2\,r_0}\approx 1.4\times 10^{-9}\ll 1$ in (\ref{M15b}).
The coordinate time difference $t_1$ is obtained by the integration
\begin{eqnarray}\label{M16a}
t_1&=&  \int_{0}^{\frac{T}{4}}\frac{dt}{d\tau}\,d\tau\stackrel{(\ref{M5},\ref{M14})}{\approx}  \int_{0}^{\frac{\pi}{2\omega}}\varphi(\tau)^{-1/2}\,d\tau\\
\nonumber
&\stackrel{(\ref{M13},\ref{M15c})}{\approx} &  \int_{0}^{\frac{\pi}{2\omega}}
\left(  1+ \frac{g}{c^2}\left(r_0+(z_1-z_0+\delta z_0)\cos  \omega \tau \right) \right)\,d\tau
\;.\\
\label{M16b}&&
\end{eqnarray}
$\delta t$ is the part of (\ref{M16b}) that is proportional to $\delta z_0$ and hence
\begin{equation}\label{M17}
 \delta t \gtrsim \int_{0}^{\frac{\pi}{2\omega}}\delta z_0\,\frac{g}{c^2}\cos \omega\tau\,d\tau=\frac{g}{c^2\,\omega}\,\delta z_0
 \;,
\end{equation}
which proves (\ref{M9}).

\subsection{Second version of Einstein's box}\label{sec:SV}

It will be instructive to discuss further possible objections to the above reconstruction of Bohr's response to Einsteins's box example.
The choice of $\tau_1=T/4$ for the proper time difference between the opening of the shutter and the (second) position measurement
of the pointer attached to the box is not the only possibility. Another choice would be  $\tau_2=T/2$ corresponding to the proper time
between two extremal values of $z(\tau)$, the minimal value $z(0)=z_0$ and the maximal one $z(\tau_2)=z_2$ such that $z_2-z_0= 2 (z_1 -z_0)$
(ignoring for the moment the  uncertainty $\delta z_0$). This alternative choice has two consequences:
\begin{enumerate}
  \item After a rotation of the harmonic oscillator's phase space with the angle $180^\circ$ the uncertainties $\delta z$ and $\delta p$ do not swap but
  assume their original values, see Figure \ref{FIGCIRC},  i.~e., (\ref {M7a}-\ref{M7b}) has to be replaced by
  \begin{eqnarray}
\label{Mixa}
  \delta z_2 &=& \delta z_0\;, \\
  \label{Mixb}
  \delta p_2 &=&  \delta p_0
  \;.
\end{eqnarray}
  This blocks the arguments leading to (\ref{M8b}) and (\ref{M10}).
  \item The integral analogous to (\ref{M17}) would vanish if extended over the interval $[0, \frac{\pi}{\omega}]$ and hence
  only the special-relativistic time dilation remains that is by a factor $m/M$ too small. Thus
  the analogous derivation of (\ref{M9}) would completely break down.
\end{enumerate}

Fortunately, both problems can be solved at the same time, by ``killing two birds with one stone":
The uncertainty of the initial position of the box system in phase space need not be of ``horizontal type'' only,
leading to an uncertainty $\delta z_0$,
but could also be of ``vertical type'' to be described by an uncertainty $\delta p_0$ of the initial momentum.
This corresponds to an uncertainty $\delta \phi$ of the initial angle variable $\phi$ in phase space
given by
\begin{equation}\label{deltaphi}
  \delta \phi \approx \arcsin \frac{\delta p_0}{M\,\omega\,(z_1-z_0)}
  \;,
\end{equation}
see Figure \ref{FIGCIRC}. Hence (\ref{M13}) has to be replaced by
\begin{equation}\label{Modi}
 r(\tau)=r_0+ z(\tau) \equiv r_0+ (z_0-z_1)\,\cos\left( \omega \tau -  \delta \phi\right)
 \;,
\end{equation}
and the integral (\ref{M16b}) by
\begin{eqnarray}\label{intmod1}
  t_2&=&\int_{0}^{\frac{\pi}{\omega}}\varphi(\tau)^{-1/2}\,d\tau\\
  \nonumber
  &\approx&
  \int_{0}^{\frac{\pi}{\omega}}\left(1+\frac{g}{c^2}(z_0-z_1)\cos\left( \omega \tau -  \delta \phi\right) \right)\,d\tau\\
  \label{intmod2}
  &&\\
  \label{intmod3}
  &=&\frac{\pi}{\omega}-\frac{2\,g\,(z_0-z_1)}{\omega\, c^2} \, \sin\delta \phi\\
  \label{intmod3}
  &\stackrel{(\ref{deltaphi})}{\approx}&  \frac{\pi}{\omega}+\frac{2\,g\,\delta p_0}{M\,\omega^2\, c^2}
  \;.
\end{eqnarray}

The latter yields
\begin{equation}\label{deltat}
 \delta t \approx \frac{2\,g}{M\,\omega^2\, c^2}\delta p_0
 \;.
\end{equation}

Next note that instead of (\ref{M4}) we now have
\begin{equation}\label{modM4}
 k(z_2-z_0)=2 g m
 \;,
\end{equation}
and hence
\begin{equation}\label{deltaEmod}
 \delta E = c^2 \delta m \approx c^2 \frac{k}{2\,g}\delta z_2 \stackrel{(\ref{Mixa})}{=} \frac{c^2\,k}{2\,g}\delta z_0
 \;.
\end{equation}
Finally,
\begin{eqnarray}\label{HUPmod}
 \delta E \,\delta t & \stackrel{(\ref{deltaEmod},\ref{deltat})}{\approx}&
\frac{ c^2 \,k}{2\,g}\delta z_0\,\frac{2\,g}{M\,\omega^2\, c^2}\delta p_0\\
 &\stackrel{(\ref{M5})}{=}& \delta z_0\,\delta p_0 \stackrel{(\ref{M6a})}{\gtrsim} h
 \;,
\end{eqnarray}
which confirms the HUP also for the second version of Einstein's box.

\section{Energy - shutter opening time uncertainty}\label{sec:EOT}

For the question of whether it is really necessary to invoke general relativity to defend the HUP,
it is appropriate to examine other solutions to the problem posed by Einstein's thought experiment.
Treder's derivation \cite{T71,BT88} of an energy-time uncertainty based on an alleged analogy
between the photon box performing damped oscillations and a metastable quantum state does not appear convincing.
Most attractive seems to me the very general proposal of P.~Busch \cite{B89} to use the Mandelstam-Tamm energy-time uncertainty relation
to derive an inequality of the following form
\begin{equation}\label{MTETUR}
\Delta E \, T_0 \,{\gtrsim} \,h
\;,
\end{equation}
where $T_0$ is the opening time of the shutter in the Einstein photon box experiment and $\Delta E$ is the standard deviation of the
box Hamiltonian $H$ w.~r.~t.~the initial (pure) state $\phi$,
 i.~e.,
\begin{equation}\label{DeltaE}
  \Delta E := \sigma(H)_\phi := \sqrt{\left\langle \phi \left| H^2 \right| \phi\right\rangle -\left\langle \phi \left| H \right| \phi\right\rangle^2 }
\end{equation}

For the convenience of the reader, we will briefly recapitulate the derivation of (\ref{MTETUR}) for the problem at hand in simplified form.
Let $T_0$ be the time required to transform the initial state $\phi$ to another state $U_{T_0}\phi$ orthogonal to $\phi$, i.~e.,
\begin{equation}\label{phitrans}
0=\left\langle \phi |U_{T_0}|\phi\right\rangle:=\left\langle \phi | \exp\left(-{\sf i}\,H\,T_0/\hbar \right)|\phi\right\rangle
\;,
\end{equation}
where $H$ is the Hamiltonian of the total quantum system representing the photon box.
Let $\langle A \rangle:=\left\langle \phi |A|\phi\right\rangle$
for each linear operator $A$
and replace $H$ by $H-\langle H \rangle$ in (\ref{phitrans}) which yields
\begin{eqnarray}\nonumber
 &&\left\langle \exp\left(-{\sf i}\,(H-\langle H \rangle)\,T_0/\hbar \right)\right\rangle\\
 \label{phitransmod}
 &=&
 \left\langle  \exp\left(-{\sf i}\,H\,T_0/\hbar \right)\right\rangle\,\exp\left({\sf i}\,\langle H \rangle T_0/\hbar \right)
 \stackrel{(\ref{phitrans})}{=}0
 \;.
\end{eqnarray}
The real part of (\ref{phitransmod}) reads
\begin{equation}\label{real}
 0=\left\langle  \cos\left((H-\langle H \rangle)\,T_0/\hbar \right)\right\rangle
 \;,
\end{equation}
which, by means of  $\cos x\ge 1-x^2/2$ and functional calculus, gives rise to the inequality
\begin{eqnarray}\label{realin1}
  &0&\ge \left\langle{\mathbbm 1}-\frac{1}{2}\left((H-\langle H \rangle)\,T_0/\hbar \right)^2 \right\rangle\\
  \label{realin2}
   &\Leftrightarrow &   \left\langle \left(H-\langle H\rangle \right)^2\right\rangle \,T_0^2 \ge 2 \hbar^2\\
   \label{realin3}
   &\stackrel{(\ref{DeltaE})}{\Leftrightarrow} & \Delta E \, T_0 \ge \sqrt{2}\,\hbar=\frac{h}{\sqrt{2}\,\pi}
   \;.
\end{eqnarray}
It can be argued that if the initial state $\phi$ represents the box with a closed shutter then the state $U_{T_0}\,\phi$
representing the box with open shutter should be approximately orthogonal to $\phi$.
(Hence the opening time would rather be $2\,T_0$ but the factor $2$ is irrelevant in (\ref{MTETUR}).)
Weakening (\ref{phitrans})  to
\begin{equation}\label{phitransweak}
  0\approx \left\langle \phi \left| U_{T_0}\right|\phi\right\rangle
  \;,
\end{equation}
then yields (\ref{MTETUR}) as a corresponding weak form of (\ref{realin3}).\\

\begin{figure}[t]
\centering
\includegraphics[width=1.0\linewidth]{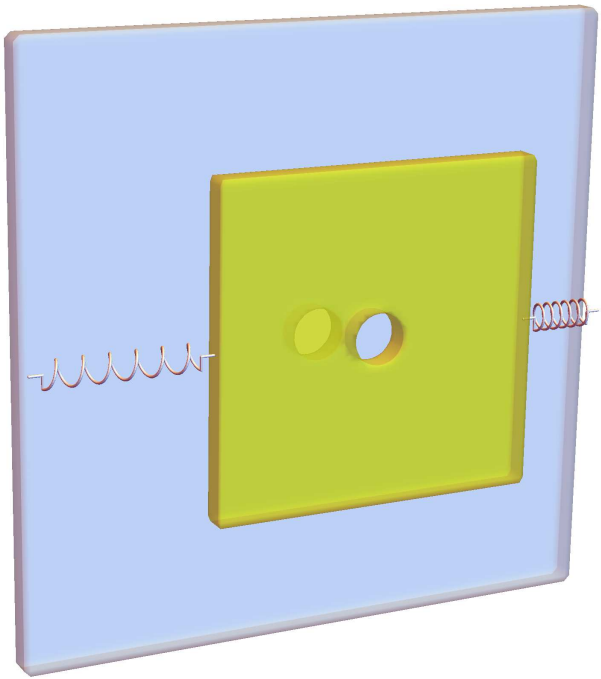}
	\caption{Sketch of a physical realization of a shutter mechanism by a vibrating
slab (yellow region) with a circular hole of diameter $L$
that overlaps with the hole in the wall of the box  (blue region) for a short moment of duration $T_0$.
}
\label{FIGSHUT}
\end{figure}

According to what has been said in Section \ref{sec:G} the implicit rules of the Einstein-Bohr debate require
that arguments using pure quantum theory have to be reformulated in the semi-classical theory considered above.
This would also hold for energy - shutter opening time uncertainty relation derived above.
Instead of the rather general form of (\ref{MTETUR}),
we will use a concrete physical implementation of the shutter mechanism by a small hole of diameter $L$ in the box wall
and a vibrating slab with a similar hole, see Figure \ref{FIGSHUT}. The vibrating slab will be modelled as a harmonic oscillator
with mass $\mu$ and total energy $\varepsilon$ and position-momentum variables  $(q,p)$ subject to the HUP
\begin{equation}\label{shutterHUP}
  \delta p\, \delta q  \,{\gtrsim} \,h
  \;.
\end{equation}
If the amplitude of the oscillation of the slab is large compared with $L$
the two holes in the wall and in the slab will overlap for approximately the time $T_0\simeq L/v$ allowing the passage of a photon,
where $v$ is the maximal velocity of the oscillating slab given by
\begin{equation}\label{vdisk}
  \frac{\mu}{2}\, v^2 = \varepsilon
  \;,
\end{equation}
and  hence satisfying
\begin{equation}\label{deltav}
  \delta \varepsilon\simeq \mu\, v\, \delta v = v\, \delta p
  \;.
\end{equation}
Due to the statistical fluctuations of the harmonic oscillator distribution the temporal position of the opening time window will be
uncertain by an amount of
\begin{equation}\label{deltaT}
  \delta t \simeq \frac{\delta q}{v}\stackrel{(\ref{shutterHUP})}{\gtrsim} \frac{h}{v\,\delta p}
  \stackrel{(\ref{deltav})}{\simeq}\frac{h}{\delta \varepsilon}
  \;,
\end{equation}
which yields the uncertainty relation
\begin{equation}\label{deltatdeltae}
 \delta \varepsilon\, \delta t \,{\gtrsim}\, h
\end{equation}
analogous to (\ref{MTETUR}). There are, however, some subtle differences: Instead of the opening time
$T_0$ itself the quantity $\delta t$ occurs in (\ref{deltatdeltae}) which represents the fluctuation of the position of the time window,
not its size. But for estimating the fluctuation of the arrival times of photons, it is just as good.
The second difference from (\ref{MTETUR}) is that $\delta \varepsilon$ is the standard deviation not of the total energy of the box,
but only of a part, namely the shutter mechanism. This is also sufficient for the present purpose
since one can with good reason exclude that the variations of $\varepsilon$ are compensated by other parts of the box
if the initial distribution is chosen in product form w.~r.~t.~variables of the shutter mechanism and the remaining part of the box.

Concerning the box experiment as modeled in Section \ref{sec:FV} one could say that validity of the energy-time HUP (\ref{deltatdeltae})
already limits the uncertainty of the {\em first} weighing before $\tau=0$. The statistical fluctuation of the total energy before $\tau=0$,
or, equivalently of the total  mass of the box,  means that we cannot longer assume that this first weighing can be performed with
arbitrary small uncertainty independent of the uncertainty $\delta t$ of the emission time of the photon.
On the other hand, the relativistic calculations in the sections \ref{sec:FV} and \ref{sec:SV},
which aim to reconstruct Bohr's response, refer to the  {\em second} weighing process at time $\tau=T/4$ (or $T/2$).

In summary, we have translated the Busch proposal \cite{B89} into the semi-classical language of the
Einstein-Bohr debate and confirmed that Bohr could have formulated a refutation of Einstein's objections
without recourse to general relativity.

\section{Third version of Einstein's box}\label{sec:TV}

\begin{figure}[t]
\centering
\includegraphics[width=1.0\linewidth]{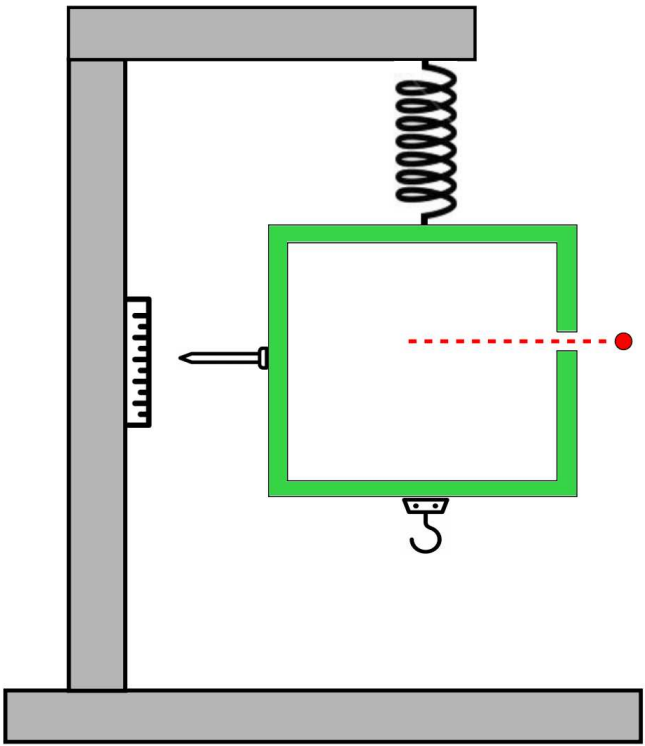}
	\caption{A schematic drawing of the third version of the photon box, drawn by deleting items from \cite{EPB}.
The box contains only one photon which eventually escapes through the hole in the wall. At the beginning the
total rest mass $M+m$ is measured by weighing; after escape of the  photon the time $t_1$ is measured at which the oscillating  box passes the new
equilibrium position.}
\label{FIGEL1}
\end{figure}

It will be instructive to continue the fictitious discussion between Einstein and Bohr for one more round.
An obvious response to the objection described in the last Section would be to dispense with a shutter mechanism altogether.
Instead, a hole in the wall of the box could be kept open all the time,
through which the photon would eventually escape and be traced in a detector.
In this detector, an energy or time-of-arrival measurement could still be taken, as desired.

Of course, the question then arises how to determine the time of emission of the photon without controlling a shutter.
In terms of classical mechanics, one can argue that this is also the time at which the box begins to oscillate around its new rest position
at $z=z_1$. Thus, if we measure the time $t_1$ at which the pointer of the box passes the plane $z=z_1$
and subtract from it the quarter period $T/4$, we obtain the time $t_0=t_1-T/4$ of the emission of the photon.
Thus, compared to the version in Section \ref{sec:FV},
we have replaced a position measurement at time $t=t_1$ with an arrival time measurement at the position $z=z_1$.
Apparently, the problem then arises that the calculation of both $z_1$ and $T$ requires knowledge of the rest mass $M$ of the box without the emitted photon.
We solve this problem by assuming that the box contains only one photon at the beginning, see Figure \ref{FIGEL1},
and that the rest mass $M$ of the empty box can either be calculated or measured independently of the actual experiment with arbitrary precision.
After all, the rest mass of an elementary particle is not an observable in the usual sense, but a physical constant.
$M$ could therefore be calculated as the sum of the masses of the elementary particles that make up the box minus the binding energy divided by $c^2$.

The latter consideration is of course not valid for the rest mass $M+m$ of the system box plus photon,
which therefore has to be measured at the beginning of the experiment by a weighing (following Einstein)
by means of a position measurement of $z_0$ (following Bohr).
We then encounter the problem of describing the ``state after the first measurement" in semiclassical theory.
In full quantum mechanics there is an elaborated theory for this, see, e.~g., \cite{BLPY16},
but it need not be adopted because of the special rules of the Einstein-Bohr debate.
Nevertheless, it seems reasonable to conceptualize the first position measurement as a
phase-space observable measuring jointly position $z_0$ and momentum $p_0$ with respective uncertainties $\delta z_0$ and $\delta p_0$
satisfying $\delta z_0 \delta p_0{\gtrsim} h$, see \cite{W04}.
The state after this measurement could then be taken as the ground state of a harmonic oscillator as far as the degree of freedom of the center-of-mass coordinates of the loaded box is concerned.
However, these considerations serve only to give the reader a sense of the physical situation; for the Einstein-Bohr debate,
it is merely assumed that the state after the first measurement satisfies the HUP.

Thus we can describe the third version of the Einstein box in two different ways: either as a preparation of an ensemble of correlated systems
``box plus photon" (3.1) or as a preparation of an ensemble of photons (3.2).
In both readings, the composite system ``box plus photon" is brought into a particular state at the beginning by an initial position measurement. After the emission of the photon, a time measurement is made at the box, which can be used to define the time zero point for possible arrival time measurements of the photon.
The said correlation refers to the energy, or rest mass, of the two subsystems in the sense that the initial rest mass $M+m$ equals the sum of the final values of the rest mass $M$ of the empty box and the energy-mass equivalent $m$ of the photon.
It is also assumed that the time $t_0$ of emission of the photon coincides with its arrival time at the detector
minus the time required for the path from the box to the detector.
In this sense, the situation is somewhat similar to the production of a correlated EPR pair,
which has already been noted and discussed by \cite{DL08} and \cite{N12}.

We focus on the second way (3.2) and interpret the uncertainties $\delta E$ and $\delta t$ as standard deviations of the energy and arrival time measurements, resp.~,
in the ensemble of photons. These standard deviations are estimated using semi-classical calculations and the correlations mentioned above.
For the first energy measurement we have the balance condition (\ref{M1})
and hence
\begin{eqnarray}
\label{deltam}
  \delta m &\approx&\frac{k}{g}\delta z_0\;,\\
  \label{deltaE}
  \delta E &\approx& \frac{c^2\,k}{g}\delta z_0\stackrel{(\ref{M5})}{=}\frac{c^2\,M\,\omega^2}{g}\delta z_0
  \;,
\end{eqnarray}
using $\delta M\approx 0$, as explained above.

The second measurement of the time $t_1$ at which the pointer of the box passes the plane $z=z_1$
gives the exact emission time $t_0=t_1-T/4$ only if the box is at rest at time $t=t_0$.
If, however, it has a certain momentum $\delta p_0$ at time $t=t_0$,
it will reach the plane $z=z_1$ slightly earlier (for $\delta p_0>0$) which leads to time uncertainty
\begin{equation}\label{deltat}
  \delta t \approx \frac{k}{M\,m\,g\,\omega^2}\delta p_0\stackrel{(\ref{M5})}{=}\frac{1}{m\,g}\delta p_0
  \;.
\end{equation}
This follows from
\begin{equation}\label{deltatphi}
 \delta t =\frac{1}{\omega}\delta \varphi
 \;,
\end{equation}
and
\begin{equation}\label{deltaphi}
\delta \varphi \approx\frac{\delta p_0}{M\,(z_1-z_0)\,\omega}\stackrel{(\ref{M4})}{=}\frac{k\,\delta p_0}{M\,m\,g\,\omega}
\;,
\end{equation}
see Figure \ref{FIGCIRC}.
Hence we obtain
\begin{eqnarray}
\label{deltaEdeltat1}
 \delta E\,\delta t &\stackrel{(\ref{deltaE},\ref{deltat})}{\approx}& \frac{c^2\,M\,\omega^2}{g}\frac{1}{m\,g}\delta p_0\,\delta z_0 \\
 \label{deltaEdeltat2}
  &\stackrel{(\ref{M6a})}{\gtrsim}& \frac{c^2\,M\,\omega^2}{m\,g^2}\,h=\left(\frac{c}{v_1}\right)^2\frac{M}{m}\,h
  \;,
\end{eqnarray}
where
\begin{equation}\label{defv1}
 v_1:=\frac{g}{\omega}\stackrel{(\ref{M5})}{=}\frac{g\,T}{2\pi}
\end{equation}
is of the order of magnitude of the velocity reached in free fall during the period of oscillation $T$ of the box.
Thus the HUP for energy and time is not only fulfilled, but is above the lower bound of $h$ by the large factor $(c/v_1)^2 (M/m)$,
see (\ref{deltaEdeltat2}).
Note that the difference between proper time of the moving box and the coordinate time that was crucial in the first two versions
of Einstein's box in the Sections \ref{sec:FV} and \ref{sec:SV} has been neglected here.
Probably the gravitational time dilation would give a further contribution
to $\delta t$, which alone would be sufficient to derive the HUP $\delta E\,\delta t\gtrsim h$, but is not needed here.

Summarizing, also the third version of Einstein's photon box does not threaten the consistency of (semi-classical) quantum mechanics
and general relativity is not needed to show this.

\section{Einstein's Box Consistency Check}\label{sec:CC}
So far, we have always assumed that the debate follows the pattern of Einstein proposing a measurement
that appears to violate the HUP and Bohr showing by a more detailed analysis of the measurement that the HUP is nevertheless satisfied.
But how about Bohr questioning the consistency of Einstein's proposal itself?
Let's play this out now for the case that Bohr doubts that one can determine the energy of a single photon by a weighing of the box.
He could thus argue that the photon exerts the same radiation pressure on the ceiling and the bottom of the box and
therefore does not contribute to the weight of the box.
Ironically, in this case {\em Einstein} would be forced to use general relativity to show that due to the gravitational red shift,
the radiation pressure on the ceiling is less than on the floor and that therefore the photon exerts a downward force upon the box.

We will calculate this force for the simple case of a purely vertical motion of a photon along a null geodesic of the Schwarzschild metric.
Alternatively, one could calculate the radiation pressure using general relativistic Maxwell theory with probably the same result.

With respect to an orthonormal frame field in $(1+1)$ dimensions the tangent vector of the null geodesic will be written as
\begin{equation}\label{tang1}
 {\mathbf k}= {\Omega \choose c\,k}
 \;,
\end{equation}
such that $E=\hbar \, \Omega$ will be the energy of the photon and $p=\hbar\, k$ its vertical momentum satisfying
\begin{equation}\label{enmom}
 \Omega = c\,k \Leftrightarrow E=c\,p
 \;.
\end{equation}
The gravitational red shift assumes the form
\begin{equation}\label{redshift}
 \Omega= \Omega_0\frac{\sqrt{1-\frac{2M_g G}{c^2 r_0}}}{\sqrt{1-\frac{2M_g G}{c^2 r}}}
 \;,
\end{equation}
compare \cite[6.3.5]{W84}, where we have identified the coordinate of the vertical direction with the $``r"$ of the Schwarzschild metric.
Let $r_0$ be the coordinate of the center of the box with length $B$ and write $r=r_0+\zeta$. Then, in linear approximation w.~r.~t.~$\zeta$, $\Omega$
can be written as
\begin{equation}\label{Omegalinear}
 \Omega\simeq\bar{\Omega} -\Omega_0\,\sqrt{1-\frac{2M_gG}{c^2 r_0}}\,\frac{g\,\zeta}{c^2}
 \;,
\end{equation}
using the gravitational acceleration $g$ according to (\ref{M13}).
Analogously,
\begin{equation}\label{plinear}
 p=\frac{\hbar}{c}\Omega\simeq\bar{p} -\hbar\,\Omega_0\,\sqrt{1-\frac{2M_gG}{c^2 r_0}}\,\frac{g\,\zeta}{c^3}
 \;.
\end{equation}
Assuming perfect reflection, the momentum transfer to the box will be $2 p$ and its difference between ceiling and the bottom
(corresponding to $\zeta=B/2$ and $\zeta=-B/2$, resp.~) amounts to
\begin{equation}\label{deltap}
\Delta\,p \simeq -2\,\hbar\,\Omega_0\,\sqrt{1-\frac{2M_gG}{c^2 r_0}}\,\frac{g\,B}{c^3}
\;.
\end{equation}
Next we consider the lapse of time between two subsequent reflections of the photon at the ceiling (or the bottom) of the box.
It will be sensible to take the proper time w.~r.~t~the center of the box which reads
\begin{equation}\label{proper}
 \Delta \tau = \sqrt{1-\frac{2M_gG}{c^2 r_0}} \,\Delta t \simeq \sqrt{1-\frac{2M_gG}{c^2 r_0}} \,\frac{2\,B}{c}
 \;,
\end{equation}
using that, in lowest order of approximation, $B/c$ is the (coordinate) time the photon need to pass a box of length $B$.
Hence for the mean force $\bar{F}$ resulting from the difference between the radiation pressure at the ceiling and the bottom of the box
we obtain
\begin{equation}\label{meanforce}
 \bar{F}=\frac{\Delta p}{\Delta \tau}\simeq -\frac{2\,\hbar\,\Omega_0\,\sqrt{1-\frac{2M_gG}{c^2 r_0}}\,\frac{g\,B}{c^3}}
 {\sqrt{1-\frac{2M_gG}{c^2 r_0}} \,\frac{2\,B}{c}}=
 -\frac{\hbar\,\Omega_0}{c^2}\,g
 \;.
\end{equation}
This means that the mean force equals the weight of an extra mass of size $m=\frac{\hbar\,\Omega_0}{c^2}$ corresponding to
the energy $E_0=\hbar\,\Omega_0$  of the photon in the center of the box.

We can interpret this result to mean that Einstein's photon box experiment can be consistently understood only within the framework of general relativity. This is hardly surprising, since the experiment is based on the special-relativistic formula $E=m c^2$ and a weighing in the gravitational field.
Thus, when Bohr, for his part, uses general relativity for his answer (which, as we have seen, he was not forced to do), this does not create a paradoxical situation, as has occasionally been speculated \cite[pp. 122 - 123]{L65}. He does not reveal a mysterious connection between two independent theories, but merely moves within the framework of Einstein's proposal.

\section{Summary}\label{sec:SUM}

The Einstein-Bohr debate and especially the discussion of the photon box has been evaluated differently over time.
Initially Bohr's account in \cite{B49} dominated and promoted the interpretation as a brilliant victory (``Einstein beaten with his own weapons").
In the last decades voices have been raised either questioning Bohr's reading of the debate or criticizing his argumentation as unclear and/or superfluous,
see \cite{T71,BT88,B89,D99,TDG00,S05,L06,DL08,N12}.
In this situation, we have tried in the present article, on the one hand, to reconstruct Bohr's answer in such a way that the original intention is clearly and understandably expressed, and, on the other hand, to discuss the suggestions of a possible answer to Einstein's objections without recourse to the general theory of relativity.
Here the implicit rules of the debate play a certain role, which we understand in such a way that the discussion is conducted in a semiclassical framework which makes a dialogue between originally ``incommensurable'' positions in the sense of T.~S.~Kuhn \cite{K70} possible at all.
In this sense, we could specifically reconstruct Busch's proposal based on the Mandelstam-Tamm energy-time uncertainty in terms of an analysis of the shutter mechanism.
Further, we considered still another version of the Einstein photon box without any shutter which can nevertheless be shown to satisfy the HUP without invoking
general relativity. While this is not proof that this will always  be the case with similar versions of the photon box, it does support the somewhat sober position that there is no secret connection between quantum theory and general relativity that would be revealed by the Einstein-Bohr debate.
Moreover, the discussion in section \ref{sec:CC} shows that the photon box, properly understood, is already a general relativistic system.

The meaning of ``uncertainty'' in the Einstein-Bohr debate allows for a certain ambiguity which, as mentioned in the Introduction, is even constitutive for the debate.
In our reconstruction  we have always understood ``uncertainty'' in the neutral sense of statistical standard deviations w.~r.~t.~an ensemble of photons.
The detailed discussion of the Einstein-Bohr thought experiment sheds particular light on the need for temporal synchronization between preparation and measurement processes, which is usually taken for granted but needs to be examined closely when the exchange of single photons is important.
There exist further interpretations of ``uncertainty'' in the sense of the difference between ideal and actual measurements that have been recently
put into a rigorous framework, see \cite{W04,BLW14,WF19}.
It is likely that these interpretations can also be applied to the Einstein-Bohr debate,
but we have refrained from doing so in order not to complicate the discussion.

\begin{acknowledgments}
I thank  Thomas Br\"ocker for stimulating discussions on the subject of this paper.
\end{acknowledgments}


\begin{thebibliography}{99}


\bibitem{B49}
N.~Bohr,
Discussion with Einstein on epistemological problems in
atomic physics,
 in {\em Albert Einstein: Philosopher-Scientist}, edited by P.~A.~Schilpp,
 Open Court, LaSalle, 1949, pp. 199 -- 241

\bibitem{EPB}
{\em Einstein's Light Box} by Prokaryotic Caspase Homolog,\\ CC BY-SA 4.0 https://creativecommons.org/licenses/by-sa,
Downloaded from: https://commons.wikimedia.org/\\wiki/File:Einstein's\_light\_box.svg

\bibitem{T71}
H.-J.~Treder, The Einstein-Bohr box experiment,
 in {\em  Perspectives in Quantum Theory}, edited by W.~Yourgrau and A.~van der Merwe,
M.I.T. Press, Cambridge, Massachusetts, and London, 1971, pp. 17 -- 24


\bibitem{BT88}
H.~H.~Borzeszkowski and H.-J.~Treder,
{\it The meaning of quantum gravity } (Reidel: Dordrecht), 1988

\bibitem{B89}
P.~Busch,
On the Energy-Time Uncertainty Relation.
Part I: Dynamical Time and Time Indeterminacy,
\textit{Found. Phys.} {\bf 20} (1),  1 -- 32 (1989)


\bibitem{D99}
D.~Dieks, The Bohr-Einstein Photon Box Debate,
 in {\em  Language, Quantum, Music}, edited by M.~L.~D.~Chiara, R.~Giuntini, and F.~Laudisa,
Synthese Library (Studies in Epistemology, Logic, Methodology, and Philosophy of Science), vol 281,
 Springer, Dordrecht 1999


\bibitem{TDG00}
A.~C.~de la Torre, A.~Daleo, and I.~Garc\'{i}a-Mata,
The photon-box Bohr-Einstein debate demythologized,
\textit{Eur. J. Phys.} {\bf 21}, 253 -- 260 (2000)


\bibitem{S05}
H.-J.~Schmidt, Einstein und die Quantentheorie,
 in {\em  Kreativit\"at, Sektionsbeitr\"age des XX. Deutschen Kongresses f\"ur Philosophie, Band 1}, edited by G.~Abel,
 Universit\"atsverlag der TU Berlin 2005, pp. 731 -- 736

\bibitem{L06}
N.~P.~Landsman,
When champions meet: Rethinking the Bohr - Einstein debate,
\textit{Stud. Hist. Philos. M. P.} {\bf 37} (1), 212 -- 242 (2006)


\bibitem{DL08}
D.~Dieks and S.~Lam,
Complementarity in the Einstein-Bohr photon box,
\textit{Am. J. Phys.} {\bf 76} (9), 838 -- 842 (2008)


\bibitem{N12}
H.~Nikoli\^{c},
EPR before EPR: a 1930 Einstein - Bohr thought experiment revisited,
\textit{Eur. J. Phys.} {\bf 33}, 1089 -- 1097 (2012)


\bibitem{W84}
R.~M.~Wald,
{\it General Relativity} (The University of Chicago Press: Chicago and London), 1984


\bibitem{BLPY16}
P.~Busch, P.~J.~Lahti, J.-P.~Pellonp\"a\"a and K.~Ylinen,
{\em Quantum Measurement},
Springer-Verlag, Berlin, 2016.


\bibitem{W04}
R.~F.~Werner,
The uncertainty relation for joint measurement of postion and momentum,
\textit{Quantum Inf. Comput.} {\bf 4} (6), 546 -- 562 (2004)

\bibitem{K70}
T.~S.~Kuhn,
{\it The structure of scientific revolutions}, 2nd ed., (The University of Chicago Press: Chicago and London), 1970


\bibitem{BLW14}
P.~Busch, P.~Lahti, and R.~F.~Werner,
Measurement uncertainty relations,
\textit{J. Math. Phys.} {\bf 55}, 042111 (2014)

\bibitem{WF19}
R.~F.~Werner and T.~Farrelly
Uncertainty from Heisenberg to Today,
\textit{Found. Phys.} {\bf 49}, 460 -- 491 (2019)

\bibitem{L65}
A.~Land\'{e},
{\it New Foundations of Quantum Mechanics},  (Cambridge University Press:
Cambridge), Massachusetts ), 1965




\end{thebibliography}
\end{document}